\documentclass{article}

\usepackage{arxiv}
\usepackage{bm}
\usepackage{natbib}
\usepackage[utf8]{inputenc} 
\usepackage[T1]{fontenc}    
\usepackage{hyperref}       
\usepackage{url}            
\usepackage{booktabs}       
\usepackage{amsfonts}       
\usepackage{nicefrac}       
\usepackage{microtype}      
\usepackage{lipsum}
\usepackage{graphicx}
\graphicspath{ {./images/} }

\title{Optimized but Unowned: How AI-Authored Goals Undermine the Motivation They Are Meant to Drive}

\author{
Vivienne Bihe Chi \\
 Computer and Information Science\\
  University of Pennsylvania\\
  Philadelphia, PA, USA \\
  \texttt{vchi@seas.upenn.edu} \\
   \And
 Roman Rietsche\\
 Institute for Digital Technology Management\\
  Bern University of Applied Sciences\\
  Bern, Switzerland \\
  \texttt{roman.rietsche@bfh.ch} \\
  \And
 Andreas G{\"o}ldi\\
  Bern University of Applied Sciences\\
  Bern, Switzerland \\
  \texttt{andreas.goeldi@bfh.ch} \\
  \AND
  Lyle Ungar \\
  Computer and Information Science\\
  University of Pennsylvania\\
    Philadelphia, PA, USA \\
  \texttt{ungar@cis.upenn.edu} \\
  \And
 Sharath Chandra Guntuku\\
Computer and Information Science\\
  University of Pennsylvania\\
   Philadelphia, PA, USA \\
  \texttt{sharathg@seas.upenn.edu} \\
}

\usepackage{bm}
\begin{document}

\maketitle
\begin{abstract}
As AI tools become embedded in productivity and self-improvement contexts, a pressing question emerges: what happens when AI does the goal-setting for us? 
In a preregistered experiment ($N=470$), we compared self-authored goals against LLM-authored goals derived from a personal reflection.
LLM-generated goals scored higher on SMART criteria ($|d|=2.26$), yet participants in the LLM condition reported lower psychological ownership ($|d|=1.38$), commitment ($|d|=1.19$), and perceived importance ($|d|=1.13$). 
At two-week follow-up, 72.8\% of self-authored participants had acted on two or more of their goals, compared to 46.6\% in the LLM condition.
Psychological ownership, not goal quality, mediated every downstream motivational outcome. 
Individuals low in trait self-efficacy, those most likely to seek AI assistance, experienced the steepest ownership erosion. 
These findings reveal a quality-motivation dissociation in AI-assisted goal-setting and identify authorship preservation as a design priority for AI tools deployed in
identity-relevant, behavior-dependent tasks.
\end{abstract}


\begin{figure}[thb]
  \includegraphics[width=\textwidth]{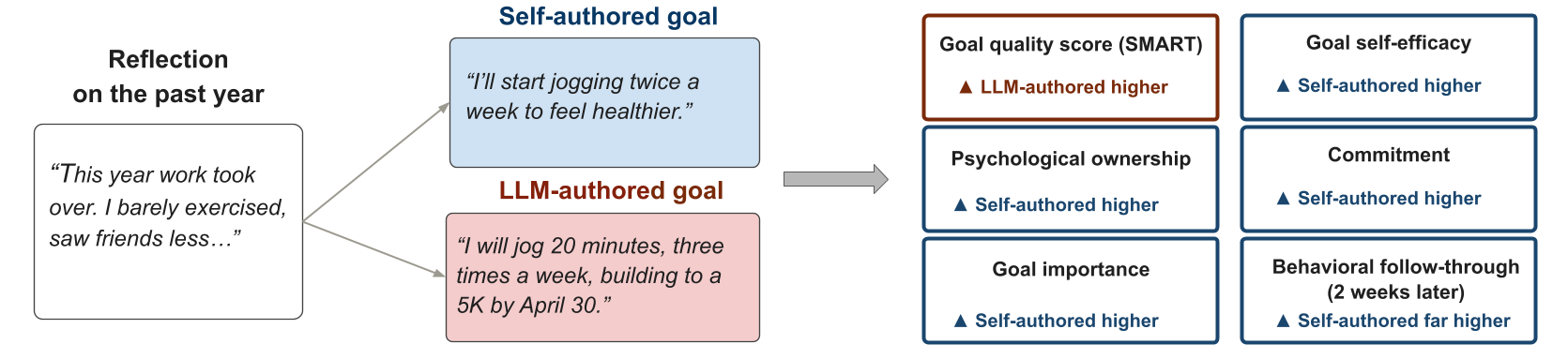}
   \caption{After a guided reflection on the past year, participants were instructed to write three SMART~\citep{doran1981} goals themselves or received LLM-authored goals based on their reflection. LLM-authored goals scored higher on objective SMART quality, yet self-authored goals produced higher psychological ownership, commitment, self-efficacy, and perceived importance, and far greater behavioral follow-through two weeks later.}
  \label{fig:teaser}
\end{figure}

\section{Introduction}
Large language models (LLMs) are rapidly being integrated into the tools people use to set and pursue personal goals. AI features in mainstream productivity platforms (e.g., Notion AI \citealp{notion_ai}; Motion \citealp{motion}) and AI coaching products (e.g., BetterUp \citealp{betterup}; Rocky.ai \citealp{rockyai}) now generate plans, surface tasks, and articulate goals on users' behalf. Within HCI, a growing body of research has examined how conversational and LLM-based systems can scaffold goal articulation, planning, and behavior change at scale \citep{mohan2020aihealthcoach,schimpf2026supporting,Bhattacharjee2024,abbas2024pitch, shah2026grow}.
The case for these systems is intuitive: decades of goal-setting research show that well-structured goals substantially outperform vague intentions in driving sustained behavior change \citep{lockeLatham2002}, and recent work suggests that AI assistance can meaningfully improve the structure of the goals users produce~\citep{schimpf2026supporting}. AI goal-setting tools could thus democratize a form of articulation expertise traditionally confined to professional coaching while reducing the cognitive load that goal-setting demands of users.

The intuitive appeal of AI-assisted goal articulation comes from a broader literature on AI writing assistance, where AI generating a clearer email, a tighter summary, or a more polished essay generally produces better outcomes for the user. Whether the same logic carries over to personal goal-setting is less obvious. In most settings where AI assistance has been studied, the artifact AI helps produce is the deliverable: the user's relationship with it ends when it is sent or submitted. A personal goal does not work this way. Its value is realized not at the moment of articulation but in the behavior it drives afterward. The artifact carries little value if the user does not act on it. Personal goals are also tied to identity in a way that emails and essays usually are not: A goal articulates who the person setting it wants to become. Recent work in HCI and communication research suggests that AI-generated content participates in shaping how users see themselves~\citep{gonzales2008identity,jakesch2019aimc,hohenstein2023ai}. Theoretical frameworks on AI-mediated communication and the algorithmic self extend this concern:  AI systems may increasingly mediate the narratives through which people interpret and construct their own identities~\citep{hancock2020aimc,joseph2025algorithmic}. When an AI articulates a personal goal, ownership of that goal matters. Self-determination theory frames this as a question of internalization: motivation is autonomous when behavior is experienced as self-endorsed \citep{deci2000what,sheldonElliot1999}. Research on psychological ownership makes a parallel argument about the artifact that it has motivational force when experienced as one's own \citep{pierce2003state}.

Even in writing-assistance contexts that lack the demands of personal goal-setting, AI assistance has been shown to lower users' sense of ownership over the resulting text \citep{draxler2024,joshiVogel2025,kadoma2024workplace}. Whether this pattern extends to personal goal-setting, and whether it has consequences for the behavior goals are intended to drive, is an open question. This paper takes up the question directly:
When an AI writes a person's personal goals,  how does AI authorship affect their motivation to pursue those goals?

To answer this question, we conducted a preregistered between-subjects experiment (N = 470) comparing self-authored and LLM-authored personal goals, with behavioral outcomes assessed at a two-week follow-up. Our findings reveal a sharp dissociation: LLM-authored goals were objectively better-formed, yet participants experienced them as substantially less their own and were less likely to act on them. 
Psychological ownership emerged as the central mediator, fully accounting for the motivational effects and partially accounting for behavior follow-through.

This paper makes three contributions to HCI. Empirically, we provide preregistered evidence that AI authorship of personal goals can simultaneously improve goal structure and undermine goal motivation. This dissociation between quality and motivation has not been captured by the goal-setting or AI-assistance literatures considered separately. 
Theoretically, we identify psychological ownership as the mechanism through which AI authorship of identity-relevant artifacts undermines motivation, establishing delegated authorship as a boundary condition for goal-setting theory's quality-performance account. Authorship shapes whether users experience the artifact as their own, and that experience drives every downstream consequence we measured.
For design, we draw out implications for AI-assisted goal-setting and related self-regulation tools, and show that users low in trait self-efficacy face the steepest motivational cost.

\section{Background and Related Work}

Two research streams bear on LLM-authored personal goals, and they look in different directions. Goal-setting theory asks what makes a goal structurally effective \citep{lockeLatham2002,lockeLatham2006}. 
Research on AI assistance asks whether AI improves outputs; its standard evaluation frameworks center on output quality, task performance, and reliance calibration~\citep{greenChen2019,bucincaTrustThink2021}.
 Neither addresses what happens when structural quality and the motivational conditions for goal pursuit come apart. Filling that gap requires two additional bodies of work:
psychological ownership and self-determination theory, which describe what makes a goal feel one's own~\citep{pierceKostovaDirks2001,ryanDeci2000, sheldonElliot1999}, and research on AI authorship in adjacent domains, which describes what changes when a model produces an identity-relevant artifact~\citep{draxler2024,joshiVogel2025,hancock2020aimc,fuTextSelf2024,riskoGilbert2016,fanMetacognitiveLaziness2025,sipondoTerblanche2026}.

\subsection{Goal Quality, Goal Source, and Goal-Setting Theory}

Goal-setting theory holds that goals influence performance through direction, effort, persistence, and strategy choice, with effects conditional on commitment, feedback, self-efficacy, ability, and task complexity \citep{lockeLatham2002,lockeLatham2006}. The SMART heuristic captures one formal slice of this theory. Doran's \citeyearpar{doran1981} original managerial formulation defined SMART objectives as specific, measurable, assignable, realistic, and time-related. Later clinical and applied variants commonly operationalize SMART goals as specific, measurable, achievable or attainable, realistic or relevant, and timed or time-bound \citep{bovendEerdt2009}. SMART is therefore a useful checklist for goal form, not a complete model of why goals work. Implementation intentions extend the operational side by binding situational cues to goal-directed actions \citep{gollwitzer1999,gollwitzerSheeran2006}, and mental contrasting combined with implementation intentions improves attainment in meta-analytic evidence \citep{wangWangGai2021}. Because commitment is itself a central condition for goal effects, goal quality should not be evaluated independently of whether the actor accepts and remains committed to the goal \citep{kleinGoalCommitment1999}. In this paper, we use \emph{formal goal quality} in this narrower sense: adherence to the five SMART dimensions. We operationalize it with a rubric-based composite score validated against independent human ratings and find that LLM-authored goals score markedly higher than self-authored goals (see Section~\ref{objective_smart_score}).

Goal source bridges this theory to AI-mediated goal formulation. \citet{lockeLatham2002,lockeLatham2006} distinguish self-set, participatively set, and assigned goals. HCI work has treated goal source as an interaction-design issue \citep{consolvoGoalSource2009,consolvoTheoryDriven2009}. In persuasive technology for physical activity, \citet{consolvoGoalSource2009} examined who should set the goal and over what timeframe, and discussed self-set, assigned, participatory, guided, and group-set arrangements. LLM authorship introduces a new source condition. The goal is not assigned by another person, jointly negotiated with an expert, or set by a peer group; an automated system generates it as a complete personal goal statement on the user's behalf. This makes authorship, not only goal specificity, central to evaluating AI goal support.

LLMs are well suited to the formal layer of goal-setting theory. They can lift a vague intention to higher specificity and measurability, attach concrete actions and time horizons, and produce plans that look correct against a SMART checklist. The closest point of comparison is \citet{schimpf2026supporting}, who tested LLM-based chatbots that operationalized goal-setting theory and implementation intentions through guidance, suggestions, and feedback. Their intervention was scaffolded goal formulation: the chatbot helped users develop higher-quality goals and implementation intentions, but did not increase goal commitment or intention to act. This makes their study strongly complementary to ours. Methodologically, they vary support features within AI-mediated goal setting; we vary the source of the final goal statement after a common reflective input and then measure ownership, recall, and action over time. Conceptually, they ask how AI can help users formulate better goals; we ask what is lost when AI authors the goal as a finished artifact. A goal that scores higher on SMART criteria but is held more weakly is not obviously a better goal \citep{lockeLatham2002,kleinGoalCommitment1999}.

\subsection{HCI Technologies for Goal Formulation and Pursuit}

HCI research has examined technologies that support both the formulation of goals and their subsequent pursuit. The questions raised by LLM authorship sit inside a longer HCI tradition of theory-driven behavior-change systems \citep{consolvoTheoryDriven2009,heklerTheoreticalGap2013}. Early persuasive-technology work proposed theory-driven design strategies for technologies that support behavior change in everyday life \citep{consolvoTheoryDriven2009}, while later HCI work argued for careful interpretation, use, and development of behavioral theory in HCI research \citep{heklerTheoreticalGap2013}. Personal informatics models further frame goal pursuit as an iterative process that involves preparation, collection, integration, reflection, and action \citep{liStageBased2010,epsteinLivedInformatics2015}. From this perspective, delegating goal formulation to an LLM is consequential because it changes the preparation stage that structures later reflection and action.

Reviews of current systems show that goal-support tools tend to implement the easier formal components of goal setting while undersupporting the adaptive components required for sustained pursuit \citep{baretta2019,ekhtiar2023,lollaSas2023}. In popular physical-activity apps, specificity and timeframes were common, whereas action planning, difficulty tailoring, and goal re-evaluation were limited or absent \citep{baretta2019}. Reviews of personal informatics and personal-goal apps similarly emphasize capture, monitoring, visualization, and reflection while calling for stronger support across goal levels, domains, and re-evaluation over time \citep{ekhtiar2023,lollaSas2023}. Reviews of behavior-change technology more broadly argue that systems often optimize engagement with the technology rather than internalization of the target behavior \citep{albertsLyngsLukoff2024}. This emphasis mirrors a broader pattern in HCI research on human--AI collaboration. In AI-assisted decision making, evaluation often centers on accuracy, reliance calibration, and overreliance; in AI-assisted writing, it often centers on writing quality, productivity, satisfaction, and ownership \citep{greenChen2019,bucincaTrustThink2021,dhillonScaffolding2024}. 
These frameworks suit artifacts whose value is realized at the moment of delivery. They are insufficient for personal goals, whose value depends on whether the user remains engaged with the output over time. A recent work on AI-assisted career-goal setting illustrates the boundary: an AI coach improved two-week progress relative to no support but not relative to a matched written-reflection questionnaire; its distinctive effect was perceived social accountability rather than self-concordance \citep{schimpfAccountability2026}. Conversational AI can move formal and short-term outcomes; whether it changes the motivational status of the goal itself is a separate question.

\subsection{Psychological Ownership and Autonomous Motivation}

Three constructs become relevant once authorship enters the picture, and they are distinct. Psychological ownership is the feeling that a target is ``mine,'' built through control, self-investment, and intimate knowledge of the target \citep{pierceKostovaDirks2001}. Autonomous motivation, in self-determination theory, is the experience of acting from one's own volition rather than under external pressure \citep{ryanDeci2000,ryanDeci2000CEP}. Self-concordance is a narrower property of the goal itself: the degree to which it expresses the person's enduring interests and values \citep{sheldonElliot1999}. The constructs come apart in practice: a goal can be self-concordant in content while feeling externally imposed in form, and a self-authored goal can feel owned even when its content is mundane.

Of the three constructs, ownership is most directly coupled to the authorship process. It assesses whether the pathways of self-investment, intimate understanding, and control were traversed at the moment of goal formulation. Autonomous motivation and self-concordance are downstream of this registration: a person is more likely to experience their behavior as self-endorsed, and to find their goals concordant with their values, when the formulation process already generated the felt sense that the goal is theirs. Ownership is therefore the predicted operative mechanism through which authorship shapes every downstream motivational consequence.

Personal goals are unusually exposed to ownership processes \citep{pierceKostovaDirks2001}. They direct attention, organize effort, and have to be enacted by the same person who holds them \citep{lockeLatham2002}. Formulating a goal is therefore not only a way of producing a textual artifact; it is one of the channels through which the control and self-investment that \citet{pierceKostovaDirks2001} identify as ownership routes can accumulate. 
The same mechanism appears in adjacent domains:
the IKEA effect shows that effort invested in assembling a product raises its
perceived value independently of quality~\citep{nortonMochonAriely2012} and the I-designed-it-myself effect in mass customization \citep{frankeSchreierKaiser2010}. 
A related effect appears directly in a goal-pursuit context: reminders recorded in one's own voice, rather than a generic alert, increase felt self-relevance and completion of daily tasks~\citep{Kim2024}.
Goal formulation may work the same way---the labor of articulating a goal builds psychological value that the resulting text alone cannot carry.

Another way of understanding goal formulation by LLM is as a form of cognitive offloading. Offloading is not inherently detrimental; external tools can reduce unnecessary cognitive demand and make difficult tasks tractable \citep{riskoGilbert2016}. The risk arises when the offloaded component is the very process through which users build understanding, monitoring, and commitment. In generative-AI-supported learning, this concern has been described as metacognitive laziness: users may receive better outputs while engaging less in the self-regulatory processes that support later understanding and action \citep{fanMetacognitiveLaziness2025, lee2025}. For goal setting, the analogous risk is not that AI produces worse goal statements, but that it displaces the reflective work through which users come to know and own their goals.

Self-efficacy enters as a theoretically important moderator. Users with lower self-efficacy may have more to gain from structured support, but commitment and persistence themselves depend on perceived capability \citep{bandura1997,lockeLatham2006}. If LLM authorship weakens ownership, structured support may carry motivational costs precisely for users whose goal pursuit is already fragile.

\subsection{AI Authorship and Scaffolded Goal Support}

AI writing provides the closest direct evidence on authorship and ownership \citep{draxler2024,joshiVogel2025}. \citet{draxler2024} report that users often did not consider themselves owners or authors of AI-generated text yet did not publicly declare AI authorship, and that greater user influence over the text increased ownership. \citet{joshiVogel2025} similarly show that AI-assisted writing lowers psychological ownership and that longer, more detailed prompts can increase it. These findings do not make goal setting a writing task; they identify a transferable authorship mechanism: ownership depends on the user's experienced contribution to the artifact. Recent HCI work on intelligent writing support converges on the same point. Support is most useful when it engages task-relevant cognitive processes rather than merely delivers finished outputs \citep{goldiWambsganssNeshaeiRietsche2024}, and the broader design space of intelligent writing assistants treats user contribution and system initiative as interaction-design choices rather than fixed properties of AI assistance \citep{leeDesignSpace2024}.

AI-mediated communication sharpens the same mechanism. In AIMC, an intelligent system operates on behalf of a communicator by modifying, augmenting, or generating messages \citep{hancock2020aimc}. Users value such tools for confidence, fluency, and expression, but also report concerns about inauthenticity and overreliance \citep{fuTextSelf2024}. Goal setting is a more self-relevant case of the same authorship problem: the generated text is not only a communicative artifact but a self-regulatory statement. Recent theorizing on the algorithmic self makes this concern broader, suggesting that AI systems increasingly mediate the narratives through which people interpret themselves \citep{joseph2025algorithmic}. This makes authorship consequential even when the generated text is formally good.

AI coaching and reflection systems provide the domain contrast. Recent reviews position AI coaching as a scalable form of structured support while emphasizing that evidence for effective design remains limited and that relational depth, cultural sensitivity, and psychological nuance remain concerns \citep{passmoreOlafssonTee2025,sipondoTerblanche2026}. Conversational agents in adjacent reflection settings show similar boundaries. In an evaluation of MindMate, a reflective-writing conversational agent, \citet{neshaeiMindMate2025} found higher writing quality, intention to use, and interactional enjoyment when students did not use the agent during the writing phase. In design-oriented work on automated coaching conversation, \citet{goldiRietsche2023} frame structured intervention and adaptive generation as alternative design directions. 
Recent work has pushed this further with LLM-based coaching frameworks that explicitly model a client's motivational readiness, tailoring emotional versus action-oriented support to their stage of change~\citep{Hoang2026}. 
The relevant design question, then, is not whether AI should be present in goal setting, but whether it supports goal formulation or replaces it.

\subsection{Research Gap}

The literature thus reveals a tension. AI can improve the formal quality of goals and implementation intentions through scaffolded formulation, and AI coaching can support short-term progress partly through felt accountability \citep{schimpf2026supporting,schimpfAccountability2026}. At the same time, psychological ownership and the autonomous-motivation family of constructs are central to whether a goal is committed to, remembered, and acted on \citep{pierceKostovaDirks2001,ryanDeci2000,sheldonElliot1999,kleinGoalCommitment1999}. 
The gap is structural. Quality research evaluates goals as artifacts to be optimized. Ownership research examines whether goals are psychologically sustained. Neither asks what happens when the two come apart. 
The missing test is authorship: whether the goal is experienced differently when the AI, rather than the user, authors the final goal statement. 

The present study isolates that test by holding the user's reflective input constant while varying the source of the final goal text.
This design separates the authorship effect from the reflection's content, making it possible to determine whether ownership routes are bypassed when the articulation step is delegated. 
If this manipulation dissociates goal quality from motivational ownership, the finding would establish a boundary condition for goal-setting theory's quality-performance account and identify a class of AI-assisted tasks for which the standard output-quality evaluation framework is insufficient.

\section{Hypotheses}
 
We test how authorship shapes the motivational consequences of personal goal setting in a two-arm, between-subjects experiment. After all participants complete the same guided reflection, those in the \textbf{self-authored} condition write their own goals from it, whereas in the \textbf{LLM-authored} condition an LLM drafts the goals from that same reflection. Three theoretical perspectives yield competing predictions about this manipulation. \emph{Psychological ownership theory} holds that ownership develops through investing the self into a target, so people should feel less ownership over what was produced \emph{for them} (e.g., a written goal) than what they produced themselves~\citep{pierceKostovaDirks2001, pierce2003state}. \emph{Self-determination theory} similarly predicts stronger motivation when behavior feels self-endorsed rather than externally caused~\citep{ryanDeci2000}. \emph{Goal-setting theory}, by contrast, suggests that well-structured goals raise perceived attainability and self-efficacy regardless of authorship~\citep{locke1990theory, lockeLatham2002}. Because LLMs are likely to produce goals that score higher on the SMART rubric than untrained authors~\citep{schimpf2026supporting}, these frameworks make offsetting predictions on different motivational outcomes.
 
We  \href{https://osf.io/cajg2/overview?view_only=847b163c96c0492cbdd9677656ed3b38}{preregistered} three confirmatory hypotheses:
\begin{description}
    \item[H1 (Commitment).] LLM-authored goals elicit lower commitment than self-authored goals, consistent with evidence that self-set goals produce stronger commitment than externally provided ones~\citep{hollenbeck1989empirical, lockeLatham2002}.
    \item[H2 (Ownership).] LLM-authored goals elicit lower psychological ownership than self-authored goals. \footnote{Prior work on AI ghostwriting has shown a small but reliable ownership decrement for AI-assisted text ($d \approx 0.21$)~\citep{draxler2024}.}
    \item[H3 (Goal Self-Efficacy).] LLM-authored goals elicit \emph{equal or higher} goal self-efficacy than self-authored goals, because clearer, more attainable goal structures raise confidence in goal pursuit~\citep{bandura1997, lockeLatham2002}.
\end{description}
 
We also preregistered three exploratory hypotheses on baseline goal perceptions:
 
\begin{description}
    \item[H4 (Importance).] LLM-authored goals are perceived as less personally important than self-authored goals, since intrinsic value derives in part from autonomous endorsement~\citep{ryanDeci2000}.
    \item[H5 (Difficulty).] LLM-authored goals are perceived as less difficult than self-authored goals, since their more structured framing raises perceived attainability~\citep{bandura1997, lockeLatham2002}.
    \item[H6 (Clarity).] LLM-authored goals are rated as clearer than self-authored goals, both subjectively and against an objective rubric~\citep{schimpf2026supporting}.
\end{description}
 
To probe behavioral follow-through, we preregistered three follow-up hypotheses\footnote{While our pre-registration included a fourth hypothesis regarding `goal importance,' we did not collect this variable in the follow-up study because it was determined to be a stable baseline measure established in the initial goal-setting session.} (Follow-up hypotheses FH1--FH3) that parallel H1, H2, and H3 at the two-week follow-up. For self-efficacy in particular, we predict that the LLM condition shows \emph{greater downward updating} from baseline to follow-up, as any initial efficacy advantage from goal structure may erode once participants engage with the underlying behavior.
 
\section{Method}
 
\subsection{Study Design}
 
\begin{figure}[hb]
 
  \centering
  \includegraphics[width=\linewidth]{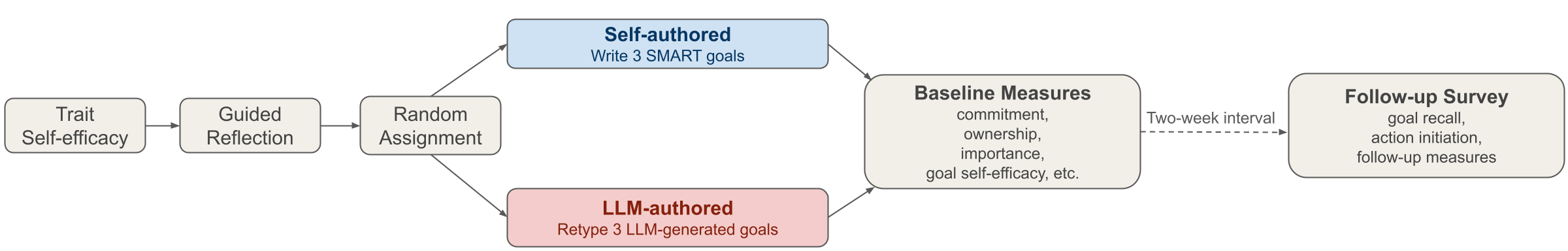}
  \caption{Study procedure. After completing a trait self-efficacy measure
  and a guided narrative reflection on the past year, participants were
  randomly assigned to a condition. Self-authored participants wrote three
  SMART goals based on their reflection. LLM-authored participants received
  three goals generated by an LLM from the same reflection and retyped them
  verbatim. Both conditions then completed identical baseline measures.
  Two weeks later, all participants returned for a follow-up survey
  assessing goal recall, action initiation, and motivational persistence.}
  \label{fig:procedure}
\end{figure}

We used a two-arm, between-subjects design (Figure~\ref{fig:procedure}) to test how goal authorship affects motivation toward personal goals. After consenting and completing baseline covariates, all participants wrote an identical guided reflection on the past year. They were then randomly assigned to draft three personal goals in one of two ways: writing the goals themselves (\textbf{self-authored}) or receiving goals drafted by an LLM from their reflection (\textbf{LLM-authored}), which they then retyped verbatim.  
The session concluded with subjective ratings of each goal and participant-level measures of ownership and perceived authorship. Two weeks later, participants returned for a follow-up survey assessing the persistence of these effects and any concrete action they had taken toward each goal. This study was reviewed by our institution's Institutional Review Board and was determined to be exempt from further review.

\subsection{Participants}
Following our \href{https://osf.io/cajg2/overview?view_only=847b163c96c0492cbdd9677656ed3b38}{preregistered} protocol, we recruited 470 U.S. adults on Prolific across the two between-subject conditions. This sample exceeds the 176 per condition required to detect $d = 0.30$ with 80\% power at $\alpha = .05$ (two-tailed) for our primary tests of H1 and H2. The effect size was informed by prior work on AI-assisted writing~\citep{draxler2024}.
 The initial goal-setting session took place in early 2026. We chose the New Year context because it provided a familiar, time-bounded setting for formulating prospective personal goals across life domains. To ensure goal setting occurred within the study rather than being recalled, we excluded participants who had already set New Year's resolutions for 2026, as assessed by a single screening item. The baseline sample had a mean age of 43.4 years ($\textit{Mdn} = 42$, $\textit{SD} = 14.1$); 53.8\% identified as women, 43.6\% as men, 1.9\% as non-binary, 0.2\% as other, and 0.4\% preferred not to say. Two weeks later, 332 participants returned for the follow-up (70.6\% retention; 169 self-authored, 163 LLM-authored).

\subsection{Measures}
 
Participants rated each of their three goals on five subjective items using a 7-point Likert scale (1 = \emph{strongly disagree}, 7 = \emph{strongly agree}): \emph{commitment} (``I am strongly committed to achieving this goal''), \emph{goal self-efficacy} (``I feel capable of accomplishing this goal in the stated time frame''), \emph{importance} (``I feel that this goal reflects what I truly want''), \emph{perceived difficulty} (``Achieving this goal would be extremely challenging for me'' ), and \emph{subjective clarity} (``I know what the goal specifically requires'').
 
Two participant-level measures asked about the three goals collectively. Psychological \emph{ownership} was assessed with four items adapted from~\citep{pierceKostovaDirks2001}, including ``I feel a very high degree of personal ownership for these goals'', ``I feel like these are \textit{my} goals'', ``I feel a strong sense of personal connection to the goals listed above'', and ``It is difficult for me to think of these goals as being \textit{mine}'' (reverse coded). As a manipulation check, participants rated \emph{perceived authorship} on a 0-100 slider (``To what extent do you feel YOU authored these goals?''; 0 = \emph{not at all authored by me}, 100 = \emph{entirely authored by me}).
 
To complement these self-reports, we also assessed each goal against an objective SMART rubric, following recent work that pairs subjective and rubric-based evaluation of LLM-supported goals~\citep{schimpf2026supporting}. Each goal was submitted to an LLM judge that independently rated each of the five SMART dimensions on a 1–5 Likert-type scale. 
We averaged the five dimension ratings to obtain a composite SMART-quality score for each goal, with higher scores indicating stronger adherence to SMART criteria.

As a robustness check on this LLM-based SMART score, we conducted a separate human-rating validation on 90 goals (45 self-authored and 45 LLM-authored), with four independent raters recruited on Prolific evaluating each goal on the same five SMART dimensions. 
Inter-rater reliability was assessed via two-way random-effects intraclass correlations (ICC(2,k), k = 4 raters).
This validation was conducted separately from the main experiment and is reported in Section \ref{objective_smart_score} as evidence of convergent validity for the rubric-based quality measure.

In the two-week follow-up, participants first performed a free-recall task, prompted with: \emph{``We're interested in what you naturally remember from the goal-setting session 2 weeks ago. It's okay if you don't remember everything exactly. Please just write what comes to mind.''}.
Afterward, we displayed the verbatim text of their original goals. For each goal, participants reported action initiation via a binary question (\emph{``In the past 2 weeks, did you take at least one concrete action toward this goal?''}, yes/no), followed by a free-response description of the action if they answered yes.

 
\subsection{Initial Goal-setting Procedure}

Participants were randomly assigned to one of the two conditions and completed the baseline session online (approximately 15 to 20 minutes). After consent, they completed the New General Self-Efficacy Scale~\citep{chen2001validation} an eight-item measure of trait self-efficacy used as a baseline moderator.  Participants were then asked to write a reflection on the past year (minimum 100 words) using goal-neutral prompts adapted from the narrative-identity framework~\citep{mcadams2001psychology}, e.g., \emph{``What were the most significant moments of the past year for you?''} (See Appendix~\ref{reflection_prompt} for the full prompt). The reflection prompts were identical across conditions, so any downstream differences are attributable to the authorship manipulation rather than to differences in self-reflective input.

Participants in the self-authored condition were given a one-paragraph definition of SMART (Specific, Measurable, Achievable, Relevant, and Time-bound) goals~\citep{doran1981} and were asked to write three goals based on their reflection. Each goal had to be 30-50 words. They could choose any life domain. The interface enforced the word range and disabled copy-paste functionality to prevent external AI use.

Participants in the LLM-authored condition were shown three LLM-generated SMART goals derived from that reflection. 
The goals were produced by a single inference call to OpenAI's GPT-5.2 using a system prompt (see Appendix \ref{llm_goal_generation_prompt} for the verbatim prompt) that constrained the output to three distinct 30-50-word SMART goals in first-person voice. The model was instructed not to include medical, legal, or crisis-related content. 
To standardize typing effort across conditions, participants were required to retype each goal into a text field verbatim. The interface validated the typed input against the generated text on a per-word basis.

For each of the three goals, participants rated goal self-efficacy, subjective goal clarity, perceived difficulty, commitment, and perceived importance in that order on a 7-point scale.
We also collected two participant-level measures: the four-item psychological ownership scale and a single-item perceived-authorship slider (0-100) as the manipulation check. 
The session concluded with demographic items (age, gender, employment status) and a single item on prior AI use frequency.

\subsection{Follow-up Study Procedure}

Two weeks after the initial session, we re-contacted all participants via Prolific to complete the follow-up survey. The follow-up took approximately 5 minutes. The survey opened with a recall task. Participants were shown three text boxes, one per goal slot, and were asked to write what they remembered about each goal from the goal-setting session two weeks earlier. The instructions explicitly allowed partial recall (``It is okay if you do not remember everything exactly. Please write what comes to mind.''). 

After the recall task, participants were shown their three original goals from two weeks ago. For each goal, they reported whether they had taken at least one concrete action toward the goal during the past two weeks (binary yes/no). Participants who answered yes described the action they took in a short free-response field.
Subsequently, participants re-completed the \emph{ownership} scale, and re-rated \emph{goal self-efficacy}, and \emph{commitment} on the same 7-point scales used at baseline. 

\subsection{Analysis}
Given unequal variances between conditions, we use Welch's $t$-tests for between-condition comparisons and report Cohen's $d$ as the effect size. Mediation and moderated-mediation models use the Hayes PROCESS macro (v4.2) with 5000-resample bias-corrected bootstraps for indirect effects, which is robust for non-normal distributions~\citep{preacher2008}. Preregistered directional hypotheses are tested one-tailed; all other comparisons are two-tailed. Composite scores are participant-level means across the three goals.

\section{Results}
\label{sec:results}
\subsection{Initial Goal-Setting Session}
\label{sec:results-baseline}

\subsubsection{Manipulation Check}

The authorship manipulation produced the intended separation. On the 0-100 perceived authorship slider, self-authored participants rated their goals as overwhelmingly their own ($M = 94.88$, $SD = 10.90$), whereas LLM-authored participants rated theirs as substantially less so ($M = 35.77$, $SD = 31.22$), $t(290.3) = 27.40$, $p < .001$, $d = 2.53$. 
The very large effect indicates that participants in both conditions experienced authorship as the manipulation intended.

\subsubsection{Rubric-based Goal Quality}
\label{objective_smart_score}
To assess whether LLM authorship produced objectively better-formed
goals, an LLM judge (Gemini 3.1 Pro; \citealp{Gilardi2023, Tornberg2024}) scored each of the 1410 goals on the five SMART criteria (Specificity, Measurability, Achievability, Relevance, Time-bound) using a structured rubric.
LLM-authored goals scored higher on every dimension and on the composite
(Self $M = 3.70$, $SD = 0.60$; LLM $M = 4.69$, $SD = 0.14$; $d = -2.26$, $p < .001$). 
By the criteria conventionally used to evaluate goal quality, LLM-authored goals were dramatically better-structured.

The separate human-rating validation further corroborated this difference. Independent human raters evaluating a subset of 90 goals on the same five dimensions reproduced the quality advantage (Self $M = 3.06$, $SD = 0.62$; LLM $M = 4.29$, $SD = 0.24$; $d = -2.60$, $p < .001$), with the advantage holding on every individual SMART dimension. Human and LLM composite ratings were strongly correlated ($r = .86$, $p < .001$), indicating that the observed quality gap was not specific to the LLM evaluator.

\subsubsection{Motivational outcomes move opposite to goal quality}
Despite the large goal quality advantage, LLM authorship produced substantially worse motivational outcomes across nearly every measure we collected.

Three preregistered predictions about motivational outcomes were strongly supported. 
Psychological ownership (H2) showed the largest condition gap. Participants in the LLM condition reported markedly lower ownership over their goals ($M = 4.19$, $SD = 1.90$) than self-authored participants ($M = 6.22$, $SD = 0.86$), $t(327.2) = -14.91$, $p < .001$, $d = 1.38$. 
Commitment (H1) followed the same pattern at comparable magnitude ($M_{LLM} = 4.53$, $SD = 1.63$ vs.\ $M_{self} = 6.05$, $SD = 0.78$; $t(335.1) = -12.94$, $p < .001$, $d = 1.19$). 
Perceived importance of the goals (H4) showed substantially lower ratings in the LLM condition as well ($M_{LLM} = 4.92$, $SD = 1.58$ vs.\ $M_{self} = 6.30$, $SD = 0.71$; $t(324.0) = -12.21$, $p < .001$, $d = 1.13$).

One preregistered prediction was reversed. 
We had predicted that LLM-authored goals would yield equal or higher goal-specific self-efficacy (H3) because they would be better structured, following goal-setting theory's proposal that clearer goals raise perceived attainability \citep{lockeLatham2002}. Instead, self-authored goals elicited higher goal self-efficacy ($d = 0.62$, $p < .001$). The structural advantage of LLM goals did not translate into greater confidence in goal pursuit.

We had also predicted that LLM-authored goals would feel less difficult (H5) and clearer (H6). While these goals were indeed rated as less difficult ($d = 0.41, p < .001$), there was no significant difference in subjective clarity ($d = -0.15$, $p = .10$), despite the LLM goals being objectively clearer by external SMART criteria.
 
The pattern across the six tests clarifies the dissociation.
Where outcomes track structural properties, LLM goals hold their own: they feel less difficult and show a marginal clarity advantage. 
Where outcomes depend on personal investment and ownership, they consistently underperform despite their structural superiority.

\begin{table*}[t]
\centering
\caption{Omnibus results for all preregistered hypotheses. Initial-session tests are based on the full sample ($N = 470$); follow-up tests are based on the retained sample ($N = 332$). Welch's $t$-tests with Mann--Whitney $U$ as robustness check}
\label{tab:omnibus}
\small
\begin{tabular}{llllrrrrl}
\toprule
\textbf{Hyp.} & \textbf{Measure} & \textbf{Prediction} & $\bm{M_{Self}}$ & $\bm{M_{LLM}}$ & \textbf{Cohen's} $\bm{d}$ & $\bm{p}$ & \textbf{Outcome} \\
\midrule
\multicolumn{8}{l}{Initial session ($N = 470$)} \\
\midrule
H1 &  Commitment (1--7) & Self $>$ LLM & 6.05 & 4.53 & $1.19$ & $<.001$ & Supported \\
H2 &  Psychological Ownership (1--7) & Self $>$ LLM & 6.22 & 4.19 & $1.38$ & $<.001$ & Supported \\
H3 &  Goal Self-Efficacy (1--7) & LLM $\geq$ Self & 5.70 & 4.98 & $0.62$ & $<.001$ & Reversed \\
H4 & Goal Importance (1--7) & Self $>$ LLM & 6.30 & 4.92 & $1.13$ & $<.001$ & Supported \\
H5 &  Perceived Difficulty (1--7) & Self $>$ LLM & 4.66 & 4.09 & $0.41$ & $<.001$ & Supported \\
H6 & Subjective Goal Clarity (1--7) & LLM $>$ Self & 6.07 & 6.18 & $-0.15$ & $.10$ & Null \\
\midrule
\multicolumn{8}{l}{Two-week follow-up ($N = 332$)} \\
\midrule
FH1 & Follow-up Commitment & Self $>$ LLM & 5.67 & 4.35 & $0.93$ & $<.001$ & Supported \\ 
FH2 &  Follow-up Ownership & Self $>$ LLM & 5.84 & 4.23 & $1.15$ & $<.001$ & Supported \\
FH3 &  Follow-up Goal Self-Efficacy & Self $>$ LLM & 5.22 & 4.55 & $0.52$ & $<.001$ & Supported \\ 
\bottomrule
\end{tabular}
\end{table*}

\subsubsection{Psychological Ownership, Not Goal Quality, Mediates the Authorship Effect}
Two competing mechanisms could in principle explain the motivational
gap. If structural quality drives motivation, the LLM advantage on the
SMART rubric should produce positive indirect effects on motivational
outcomes. If psychological ownership drives motivation, the loss of
ownership in the LLM condition should mediate the negative effects.
We tested both using parallel single-mediator models (Hayes Process Model 4).

\begin{figure}[h]
    \centering
    \includegraphics[trim={0cm 6cm 0cm 6cm},width=.98\linewidth]{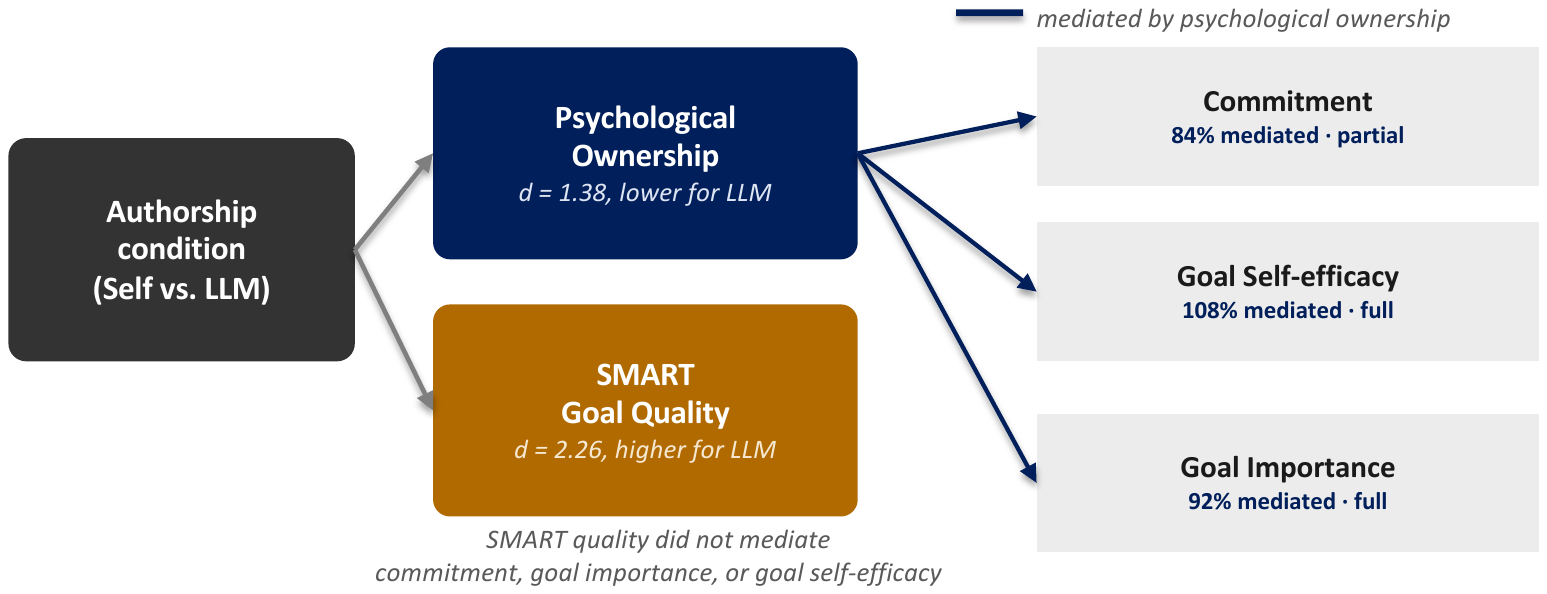}
     \caption{Psychological ownership, not SMART goal quality, mediates the effect of goal authorship on personal motivational outcomes. Single-mediator bootstrap models (5,000 resamples, bias-corrected 95\% CIs) testing authorship condition (Self vs. LLM) on commitment, goal self-efficacy, and goal importance. Ownership significantly mediated all three (84–108\%), with full mediation (a non-significant direct effect once ownership was controlled) for self-efficacy and importance, and partial mediation for commitment. SMART goal quality, despite the large LLM advantage in formal structure ($d = 2.26$), did not mediate any of the three.}
  \label{fig:mediation}
\end{figure}

Ownership significantly mediated the authorship condition effect on every downstream motivational outcome (commitment, importance, goal self-efficacy, perceived difficulty, subjective clarity).
The indirect effects were large and the proportions mediated were high (84--108\%), Figure~\ref{fig:mediation}.
For goal self-efficacy and importance, the direct effect of authorship condition became non-significant once ownership was controlled (full mediation), meaning the LLM disadvantage on these outcomes is entirely accounted for by the loss of ownership. 

The rubric-based SMART score, by contrast, did not mediate ownership, commitment, importance, or goal self-efficacy. Psychological ownership was the operative mechanism for every personal, self-referential outcome. SMART quality did shape text-level judgments, accounting for 83.5\% of the difference in perceived difficulty. This dissociation between formal and personal outcomes explains the H6 null as well: subjective clarity tracks the person's relationship to the goal, not the syntactic specificity of the text.

\subsubsection{The Ownership Deficit Is Largest Among Low-Self-Efficacy Users}
\label{sec:results-initial-moderation}

We tested whether trait self-efficacy moderates the LLM disadvantage on ownership, commitment, and goal-specific self-efficacy. Moderation was significant for ownership ($b = 0.62$, $p < .001$, $\Delta R^2 = .015$) and for commitment ($b = 0.34$, $p = .03$, $\Delta R^2 = .007$), but not for goal-specific self-efficacy ($p = .53$). Users with higher trait self-efficacy showed a smaller LLM disadvantage on ownership and commitment.
The split is informative. A general sense of self-efficacy helps users
feel that an AI-written goal is still theirs, but it does not help
them feel capable of pursuing a goal they did not formulate.

A moderated mediation model (Authorship condition $\rightarrow$ Ownership $\rightarrow$ Commitment, with trait self-efficacy moderating the $a$-path; Hayes Model~7 \citealp{preacher2008}) gave a significant index of moderated mediation (IMM $= 0.39$, 95\% CI $[0.15, 0.62]$). Conditional indirect effects were significant at every level examined: at one standard deviation below the mean of trait self-efficacy, $ab = -1.57$, 95\% CI $[-1.87, -1.27]$; at the mean, $ab = -1.29$, 95\% CI $[-1.52, -1.08]$; at one standard deviation above, $ab = -1.01$, 95\% CI $[-1.29, -0.78]$.
 
The pattern describes the magnitude of an effect that is present across the full range of self-efficacy. 
Ownership erosion was approximately 55\% larger for users low in trait self-efficacy than for users high in trait self-efficacy, but it was substantial for both. This pattern carries direct design relevance. Users low in trait self-efficacy are, on existing evidence \citep{bandura1997, schunk2020selfefficacy}, the population most likely to seek AI assistance in articulating their goals. The same users who stand to benefit most from external scaffolding are also the users for whom delegated authorship exacts the steepest psychological cost.

\subsection{Two-Week Follow-Up}
\label{sec:results-followup}

Two weeks after the baseline session, $332$ of $470$ participants returned (70.6\%; $169$ self-authored, $163$ LLM-authored). Retention did not differ by condition, $\chi^2(1) = 0.11$, $p = .74$. Returners did not differ from non-returners on baseline ownership, commitment, or trait self-efficacy (all $|d| < 0.10$, all $p > .35$).
\subsubsection{Goal recall and behavioral action}
We measured goal pursuit in two ways. 
First, we assessed free-recall accuracy of the goals set two weeks earlier.
To assess free-recall accuracy two weeks after the study, we used GPT-4o with a structured rubric to score each response against the original goal. Each was evaluated for topic accuracy (correct domain or aim) and component accuracy (specific elements such as quantities, timeframes, and actions).
Self-condition participants recalled their goals more accurately on
both measures. Independent-samples $t$-tests confirmed both differences at $p < .001$.
The gap is notable given that LLM-authored goals contained more surface-level memory cues---numbers, deadlines, and concrete actions. Structural richness did not compensate for the absence of formulation work. The process of writing a goal appears to support its encoding as well as its motivation.

\begin{figure}
    \centering
    \includegraphics[width=0.7\linewidth]{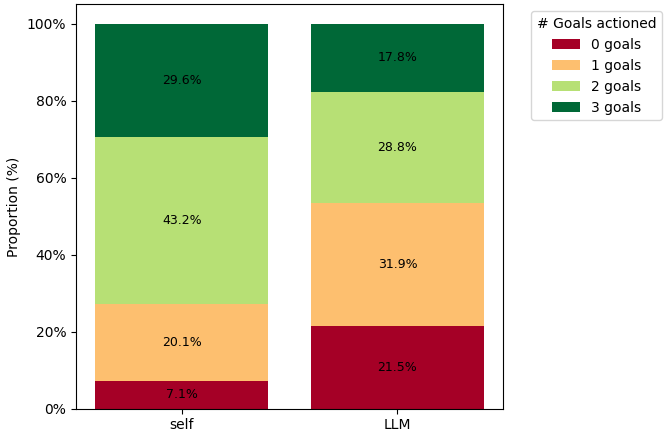}
     \caption{Distribution of goal action counts at two-week follow-up by
  condition ($N = 332$). Each bar shows the proportion of participants who
  took action on 0, 1, 2, or 3 of their goals in the two weeks following
  the initial session. Self-authored participants were substantially more
  likely to act on multiple goals: 72.8\% acted on two or more, compared
  to 46.6\% in the LLM-authored condition ($\chi^2(1) = 22.93$,
  $p < .001$).}
  \label{fig:action-initiation}
\end{figure}

Second, we recorded self-reported action: for each goal, whether the participant had taken at least one concrete action toward each goal in the past two weeks.
Self-condition participants acted on more goals ($M = 1.95$ of~3) than LLM-condition participants ($M = 1.43$), $t(327.7) = 4.99$, $p < .001$. The gap was consistent across all three individual goals (all $p < .001$) and held at the aggregate level.
At follow-up, 72.8\% of self-authoring participants had acted on two or more goals, significantly outperforming the LLM condition (46.6\%; $\chi^2(1) = 22.93$, $p < .001$; Figure~\ref{fig:action-initiation}).
The motivational gap observed at baseline thus translated into a substantial behavioral gap two weeks later.

\subsubsection{Ownership Mediates Action Initiation}
Action initiation lets us ask the strongest version of the mechanism question: does ownership predict behavior, beyond self-reports? 
In a logistic mediation with bootstrapped indirect effects ($N = 332$), baseline psychological ownership significantly mediated the effect of authorship on action initiation ($ab = 0.41$, 95\% CI $[0.02, 0.81]$), accounting for 32\% of the total effect. The direct effect remained significant once ownership was controlled ($b = 0.84$, $p = .047$), indicating partial mediation. Crucially, baseline ownership was measured at the initial goal-setting session, before participants had any opportunity to act. Therefore, ownership predicts later behavior rather than merely correlating with other self-reports.

\subsubsection{Persistence of Motivational Differences}
We next tested whether the motivational differences observed at baseline remained at follow-up.

Self-authored participants reported higher commitment to their goals at follow-up (FH1; $M_{self} = 5.67$ vs.\ $M_{LLM} = 4.35$, $t(323.0) = 8.47$, $p < .001$).
They also reported higher psychological ownership of the goals (FH2; $M_{self} = 5.84$ vs.\ $M_{LLM} = 4.23$, $t(285.8) = 10.48$, $p < .001$), and higher goal self-efficacy (FH3; $M_{self} = 5.22$ vs.\ $M_{LLM} = 4.55$, $t(305.1) = 4.69$, $p < .001$). 
All three follow-up hypotheses were supported.

Both conditions showed the expected drop from baseline to follow-up. Goal self-efficacy declined by similar amounts in both conditions (self: $-0.52$; LLM: $-0.39$; one-tailed $p = .84$ for the preregistered prediction that LLM would decline more).
We had hypothesized that LLM-condition participants would update goal self-efficacy downward more than self-condition participants once they encountered the demands of the underlying behavior; that prediction was not supported. 
Commitment also declined in both conditions, with a larger drop in the self condition than in the LLM condition ($-0.41$ vs. $-0.05$; two-tailed $p = .008$). The smaller drop in the LLM condition reflects a floor effect: LLM commitment was already low at baseline.
The motivational shortfall under LLM authorship was therefore immediate and durable rather than progressive: it was established at the moment of authorship and held steady across the two-week window.

\section{Discussion}
\label{sec:discussion}

LLMs can write better-structured goals than people typically write for themselves. Our LLM-generated goals scored over two standard deviations higher on SMART criteria. They were more specific, more measurable, and more time-bound.
However, the higher goal quality did not translate to motivational benefits. Compared to self-authored participants, those in the LLM condition reported less ownership over their goals, less commitment, and less personally important. 
Two weeks later, they recalled fewer goal components. They were far less likely to take action: only 46.6\% of LLM-condition participants had acted on more than one goal, compared to 72.8\% of those in the self-authored condition.

Three further findings extend this dissociation. Psychological ownership mediated every motivational outcome we measured. Ownership also predicted behavioral action two weeks later, beyond what authorship condition alone could explain. The ownership erosion was steepest for users low in trait self-efficacy, the population most likely to seek AI assistance in the first place.

\subsection{Why quality and ownership come apart}
\label{sec:disc-mechanism}
The standard logic of AI assistance is that objectively superior outputs provide greater utility, regardless of whether AI or the user themselves produced the output.

This logic does not apply to personal goals. Two theoretical traditions help explain why. Psychological ownership theory~\citep{pierce2003state} identifies three routes through which people come to feel something is ``theirs": investment of self, intimate knowing, and control over the target. Authoring a goal walks through all three routes at once. The user thinks about what matters, decides what to commit to, and produces the words. Delegating goal authorship to an LLM walks through none of these routes. The user invests no thinking time, gains no intimate knowledge of why one goal was selected over another, and exercises no control over the articulation. Even when the resulting goal is excellent, the work that builds ownership was not done.

Self-determination theory (SDT) makes a parallel point~\citep{deci2000what, sheldonElliot1999}. Autonomous motivation requires that behavior be experienced as self-endorsed. SDT does not require that goals originate from the self. An externally introduced goal can become autonomous through internalization. Our results suggest that, under conditions of full LLM authorship, internalization mostly does not happen within the two-week timeframe we measured. Participants treated LLM-authored goals as something received, not something internalized.

Together, these two accounts 
imply that goal quality and goal ownership operate on different psychological registers. The features that make a goal structurally good do not automatically build the features that make a person commit to it. Our mediation analyses confirm this directly. Psychological ownership accounted for the motivational gap on every personal outcome. SMART quality accounted for none of them, though it did shape text-level judgments like perceived difficulty. This dissociation is our theoretical contribution: LLM authorship can improve formal goal quality while eroding the psychological ownership that sustains motivation.


\subsection{Ownership predicts later behavior}
\label{sec:disc-behavior}
Self-report motivational measures are often subject to skepticism due to social desirability bias, and the explicit framing of goals as AI-generated may have further primed lower ownership responses. Our follow-up data address this concern by demonstrating how baseline ownership, measured at the moment of goal authorship, predicted whether participants took concrete action on their goals two weeks later. 
Mediation analysis revealed that ownership accounted for 32\% of the total effect of authorship on follow-up action ($ab = 0.41$), while the direct effect remained significant, indicating partial mediation.
This result suggests that psychological ownership is not just a transient feeling; it can predict later action.

\subsection{Connections to AI writing assistance research}
\label{sec:disc-writing}

Recent HCI work has documented similar patterns of ownership erosion in writing contexts. 
\citet{draxler2024} showed that users disclaim ownership of AI-generated text even while accepting authorship credit. 
\citet{joshiVogel2025} found that longer prompts (more user investment) preserve more ownership when writing creative fiction with AI. 
\citet{kadoma2024workplace} showed that perceived control mitigates ownership loss in workplace AI-mediated communication.
Our findings sit alongside this body of work. Ownership erodes when users delegate to AI. What our study adds is the behavioral cost. 
Drafting an email is finished when it is sent. The cost of low ownership there is mostly social. Setting a goal is only beginning at the moment of articulation. The cost of low ownership shows up in everything that follows.

Goal-setting is also identity-relevant in a way that emails usually are not. A goal articulates who the user wants to become. 
Recent work on AI-mediated communication~\citep{hancock2020aimc} and the algorithmic self~\citep{joseph2025algorithmic} suggests that AI-generated content participates in shaping how users see themselves. 
Our results give that prospect empirical traction. LLM-authored goals felt less difficult and scored higher on SMART criteria. These structural advantages did not carry over to how participants related to the goals. Lower ownership, lower commitment, and lower importance persisted regardless of the quality advantage.

\subsection{Goal-setting theory assumes self-authorship}
\label{sec:disc-goal-setting}
Classical goal-setting theory~\citep{lockeLatham2002} has been an important framework in HCI literature on behavior change for decades. 
SMART criteria, specificity-difficulty effects, and goal-commitment models were all developed under the implicit assumption that goals are self-set. The theory says little about what happens when an external system writes the goal for the user.

Our results suggest this assumption was critical to the overall effect.  The structural properties Locke and Latham ~\citep{lockeLatham2002} identified predict performance \emph{conditional on} the goal being self-endorsed. When self-endorsement is removed, the structural advantages do not carry over into behavior.

We see this as a boundary condition on Locke and Latham's ~\citep{lockeLatham2002} specificity principle. In personal domains, where the person setting the goal has rich self-knowledge that no external source can access, specificity from an outside source carries no informational value the person did not already have. The structural improvement is real but psychologically empty. Specificity translates into motivation only when it tells the user something they did not already know.

This connects to recent HCI work on LLM-supported goal-setting~\citep{schimpf2026supporting, mohan2020aihealthcoach, Bhattacharjee2024}, which has demonstrated that LLMs and conversational agents can improve goal articulation against quality benchmarks. 
Our results extend prior research by showing that quality alone cannot sustain motivation.
Goal quality is a necessary but insufficient condition for pursuit, as the motivational benefits of goal-setting depend equally on psychological ownership. AI tools that optimize purely for SMART quality may be optimizing the wrong thing.

\subsection{Design implications}
\label{sec:disc-design}

Our findings have direct implications for the design of AI-assisted goal-setting tools, and for AI assistance with any task where the output is meant to drive future behavior. Table~\ref{tab:design-principles} summarizes four design principles derived from these findings; the following paragraphs elaborate them.

\begin{table*}[t]
\centering
\caption{Design principles derived from the empirical findings.}
\label{tab:design-principles}
\small
\begin{tabular}{
    p{0.21\textwidth}
    p{0.19\textwidth}
    p{0.21\textwidth}
    p{0.28\textwidth}}
\toprule
\textbf{Empirical finding} &
\textbf{Design requirement} &
\textbf{Design principle} &
\textbf{Example instantiation} \\
\midrule

LLM authorship lowered psychological ownership. &
Preserve self-investment and control. &
Scaffolded goal articulation. &
Guide users through the SMART dimensions while keeping the goal user-authored. \\

SMART quality did not mediate motivational outcomes. &
Improve formal goal quality without replacing authorship. &
Refinement over generation. &
Begin with a user-authored draft and let the LLM suggest revisions. \\

Baseline psychological ownership predicted action initiation. &
Detect low ownership before goal pursuit. &
Ownership as a UX metric. &
Ask ``Does this feel like your goal?'' and route low-ownership users to collaborative revision. \\
The ownership deficit was largest at low trait self-efficacy. &
Provide support without doing the articulating. &
Tailoring for low trait self-efficacy. &
Increase prompts, examples, and choices while preserving user articulation. \\

\bottomrule
\end{tabular}
\end{table*}

\textbf{Scaffolded goal articulation.} Tools should help the user articulate their own goals rather than produce articulations on the user's behalf.~\citet{joshiVogel2025} showed this preserves ownership in creative writing. We extend this work by isolating authorship and showing why such support should preserve user authorship: delegated authorship lowers psychological ownership and action initiation. 

\textbf{Refinement over generation.} An AI that edits a user's draft preserves authorship. An AI that produces the draft does not. The final text can look similar in both cases, yet the psychological difference comes from where the user invested. All three of the ownership routes run through that first draft.

\textbf{Ownership as a UX metric.} Standard usability metrics miss the motivational consequences of delegated authorship. Task completion, efficiency, and satisfaction are necessary, but a goal-setting interface that ranks high on all three can still leave users disengaged. We should measure ownership directly during evaluation. 

\textbf{Tailoring intervention for low trait self-efficacy users.} The intuitive response to a user who feels uncertain about goal articulation is to do more of the work for them. Our moderation result shows this is the wrong move. The harder design problem is to scaffold articulation for users who feel uncertain, without doing the articulating for them.

\subsection{AI assistance in identity-relevant domains}

The dissociation we document is unlikely to be specific to personal goals. Wherever AI assistance is deployed in domains that depend on the user's continued engagement with the output, the same logic should apply. Therapy and counseling, educational planning, creative writing, career articulation, and reflective writing all share two properties with goal-setting. The artifact is activating rather than terminal, and it is tied to identity.

Each of these domains has seen recent investment in AI-assisted tools. The standard evaluation framework, drawn from task-completion contexts, asks whether the AI produces better outputs. Our results suggest a different question. What is the motivational cost of producing those outputs through delegation, and for whom is the cost largest? Tools that look successful by the output-quality metric may be failing by the engagement metric. The framework for evaluating AI assistance in identity-relevant domains needs to make room for both.

\subsection{Limitations and Future Work}
\label{sec:disc-limitations}

In our study, participants in the LLM condition received a complete first draft of three goals and retyped them verbatim, 
a procedure that controls for typing effort and arguably biases against rather than inflates our hypothesized ownership deficit. This isolates the authorship effect cleanly, but it tests only one point in a larger design space, as real-world AI goal-setting tools  often involve iterative, conversational refinement. Whether co-authorship designs preserve ownership while still capturing AI's structural-quality benefits is an open question.  Future work should test other partial-scaffolding designs.

We mainly measured psychological outcomes and self-reported action, not objective behavioral attainment. Our findings address the demand-characteristic critique most directly through Section~\ref{sec:disc-behavior}, where baseline ownership predicts later action with the causal direction unambiguous. Even so, real-world goal pursuit involves cycles of attempt and revision that the two-week window does not capture.
Future research should employ longer longitudinal designs to determine whether the ownership deficit persists as users disengage, or whether users gradually internalize LLM-authored goals over time.

We measured goal quality through rubric-based 
SMART criteria. Goals could also be evaluated for personal fit, or alignment with norms and values. Whether AI authorship would still produce a quality advantage on these alternative measures is unclear.

Beyond these limitations, three broader directions follow from our findings. (1) Personalization. Our results identify the authorship cost as largest when the LLM lacks access to the user's self-knowledge. AI goal assistants that incorporate personal history, values, and constraints may produce goals that feel more self-authored, narrowing the ownership gap without giving up structural quality. (2) Mechanism. Studies using think-aloud or experience-sampling methods could map when users revise, abandon, or re-engage with goals during pursuit, and whether the trajectories differ by authorship. (3) Generalization. The autonomy cost we document should be tested in adjacent identity-relevant domains where AI delegation is already common practice.

\section{Conclusion}
\label{sec:conclusion}

A widely-held assumption in AI assistance is that better outputs are better for users. 
In personal goal-setting, this assumption breaks down. AI can write objectively better goals than people write for themselves, yet people do not own and act on those goals.  Our findings identify psychological ownership as a critical driver of motivation.  Authoring a goal provides a motivational force that AI cannot replicate without high cost. In our study, this cost was substantial: a 26\% gap in behavioral follow-through, and most pronounced among users with low self-efficacy. This creates a troubling paradox: the users most likely to seek AI assistance are also the most susceptible to its ownership-diminishing effects.

These results indicate that AI must scaffold authorship rather than replace it. 
Designing for tasks that depend on continued user engagement is fundamentally different from designing for final output delivery.
The field's default criterion for evaluating AI assistance is output quality. Our findings suggest that in identity-relevant tasks, authorship preservation is as consequential as output quality.

\bibliographystyle{ACM-Reference-Format}
\bibliography{goalsetting-ref,background_proofread}

\appendix
\section{Reflection guide prompt}
\label{reflection_prompt}
\begin{verbatim}
Looking back on the past year–
Where did you spend your time, energy, and attention?
What felt rewarding, what felt frustrating in your life?

Please describe your reflections in detail, using specific examples where helpful.
Your response should be at least 100 words.
\end{verbatim}

\section{LLM goal generation prompt}
\label{llm_goal_generation_prompt}
\begin{verbatim}
You are a goal-setting coach.
Task: Based on the user’s reflection on the past year, generate exactly THREE distinct SMART goals 
        from the user's reflection.
Constraints:
- Return strictly valid JSON (no markdown) with keys: goal1, goal2, goal3.  
Each goal must be 30 to 50 words.
- Each goal must be Specific, Measurable, Achievable, Relevant, and Time-bound.
- The goals may come from any area of the user’s life (e.g., health, relationships, work, hobbies).  
- Write the goals in the first-person voice, as if the user wrote them themselves. 
- Use plain, natural language (avoid corporate tone).
- keep the goals safe; do not include medical, legal, or crisis hotlines. 
- Use only plain ASCII characters (no fancy quotes, no em dashes).
- Make the three goals clearly different (different domains or themes).
User reflection (verbatim): 
\end{verbatim}

\end{document}